\documentclass{article}

\usepackage{PRIMEarxiv}

\usepackage[utf8]{inputenc} 
\usepackage[T1]{fontenc}
\usepackage{ragged2e}
\usepackage{url}   
\usepackage{bm}
\usepackage{multicol}
\usepackage{sectsty}
\usepackage{nccmath}
\usepackage{float}
\usepackage{commath}
\usepackage{amsmath}
\usepackage{amsfonts}
\usepackage{amssymb}
\usepackage[numbers]{natbib}
\usepackage{dcolumn}
\usepackage{mathtools}
\usepackage{float}
\usepackage{fancyhdr}
\usepackage{fnpos}
\usepackage[english]{babel}
\usepackage{lastpage}
\usepackage{booktabs}
\usepackage{amsfonts}
\usepackage{mathptmx}
\usepackage{mathtools}
\usepackage{amsfonts}

\makeatletter
\def\blfootnote{\xdef\@thefnmark{}\@footnotetext}
\makeatother
\usepackage{nicefrac}
\usepackage{microtype}
\usepackage{lipsum}
\usepackage{fancyhdr}
\usepackage{graphicx} 
\graphicspath{{media/}}

\title{de Broglie-Bohm  analysis of a nonlinear membrane: From quantum to classical chaos}
 \author{Henrique Santos Lima\textsuperscript{1},  Matheus M. A. Paixão\textsuperscript{1}\hspace{0.5mm} and Constantino Tsallis\textsuperscript{1,2,3}}

\date{\today}
\begin{document}
\maketitle

\begin{abstract}
Within the de Broglie-Bohm theory, we numerically study a generic two-dimensional anharmonic oscillator including cubic and quartic interactions. Our analysis of the quantum velocity fields and trajectories reveals the emergence of dynamical vortices. In their vicinity, fingerprints of chaotic behavior such as unpredictability and sensitivity to initial conditions are detected. The simultaneous presence of off-diagonal and nonlinear terms leads to robust quantum chaos very analogous to its classical version.
\end{abstract}
\blfootnote{\textit{\textsuperscript{1}~Centro Brasileiro de Pesquisas Fisicas, Rua Xavier Sigaud 150, Rio de Janeiro-RJ 22290-180, Brazil \\
E-mail: hslima94@cbpf.br/matheuspaixao@cbpf.br}}
\blfootnote{\textit{\textsuperscript{1,2,3}~Centro Brasileiro de Pesquisas Fisicas and National Institute of Science and Technology of Complex Systems, Rua Xavier Sigaud 150, Rio de Janeiro-RJ 22290-180, Brazil \\
Santa Fe Institute, 1399 Hyde Park Road, Santa Fe, 
 New Mexico 87501, USA \\
Complexity Science Hub Vienna, Josefst\"adter Strasse 
 39, 1080 Vienna, Austria \\
E-mail: tsallis@cbpf.br}}
\section{Introduction}
 Quantum chaos~\cite{Peres1984, Izrailev1990,gutzwiller1990} is a rich problem in the realm of quantum mechanics. Moreover, it has a wide variety of applications, e.g.,  quantum computing ~\cite{Georgeot2000,Shepelyansky2001,Georgeot2001,Prosen2001}, quantum dots \cite{Luscher2001,Ponomarenko2008}, nuclear physics~\cite{Bohigas1988,Mitchell2010} and even in cosmology and black holes~\cite{Magan2018}. Our present goal for advancing the understanding of this phenomenon is to scrutinize the correspondence between quantum and classical chaotic dynamics by tunning the value of the Planck constant.  Starting from classical dynamics, it is necessary to integrate all equations of motion to determine whether a system is dissipative, ordered, strongly  or weakly  chaotic. 
Conversely,  in quantum mechanics, the Schrödinger equation $\mathcal{H}\Psi=i\hbar\partial_t \Psi$ is linear and acts on probability amplitudes instead of trajectories. So, to study quantum chaotic dynamics, many authors suggest that the de Broglie-Bohm (dBB) theory ~\cite{deBroglie,Bohm1,Bohm2} is an intriguing  alternative to investigate quantum chaos ~\cite{Parmenter1995,Faisal1995,dePolavieja1996,Wu1999,Cushing1999,Cushing2000,Conto2006,Contopoulos2008a,Borondo2009,Sengupta2014,Conto2020,Tzemos2020,TzemosContopoulos20222,Tzemos2022}, as it is based on trajectories determined  by heavily nonlinear terms. On classical grounds, the Liouville equation 
\begin{equation}
  \partial_t \rho=-\{\rho,\mathcal{H}\}\equiv-i\mathcal{L}\rho,
    \label{Liou}
\end{equation}
where  $\rho$ is the probability density, $\{ .\}$ is the Poisson bracket and $\mathcal{L}$ is the Liouvillian operator, governs the time evolution of the system.  From Eq. \eqref{Liou} it is not possible to directly  measure  chaos or any consequence of it since, like the Schrödinger equation, it is linear. Quantum observables operate  in the Hilbert space in such a way that the corresponding distributions are typically well-behaved, thus apparently suggesting that there is no chaos in quantum mechanics. This is obviously inadmissible since quantum mechanics recovers classical mechanics (widely known to exhibit chaotic behavior) in the $\hbar\to0$ limit; examples of classical chaos are found in~\cite{Lecar2001,Ma2017,Bo2021,Mihailovic2014}. Although Eq.  \eqref{Liou} is linear, a classical system with more than three equations of motion and nonlinearities is likely to exhibit sensitivity to initial conditions, hence some type of chaos.  Therefore, at least in classical mechanics, the feasible way to analyse the collective behavior of particles is via equations of motion. 
In this sense, in the Bohmian interpretation of quantum mechanics,  nonlinear effects naturally arise from the equations of motion due the additional presence of the quantum potential $Q\equiv -\frac{\hbar^2}{2m}\frac{\nabla^2 |\Psi|}{|\Psi|}$, which exhibits the equivalence between quantum and classical trajectories in the limit  $\hbar \to 0$. Conversely, when the quantum potential diverges in two and three-dimensional systems, it can result in the emergence of vortices~\cite{Frisk1997,Birula2000,Wisniacki2005,Wisniacki2006,Wisniacki2007}, an important component in the study of superconductors~\cite{Blatter1994}, Bose-Einstein condensates~\cite{Fetter2009,Solnyshkov2019}, superfluid phenomena~\cite{Ruutu1996,Autti2016}, quantum field theory~\cite{Rajaraman1975,Frohlich1989,DeLima2022}, to cite but a few. The appearance of these quantum vortices is an important element in the emergence of chaos due to the presence of  geometric structures called nodal point X-point complexes (NPXPCs)~\cite{Conto2009,Falsaperla2003,Tzemos2020}. Additionally, chaotic dynamics constitutes a crucial factor in validating Born's rule for systems that do not satisfy the quantum equilibrium hypothesis~\cite{Valentini1991I,Valentini1991II,Conto2021}, i.e., for initial conditions that are not distributed according to $|\Psi|^2$. 

Here, we conduct a full numerical study of a generic two-dimensional quantum anharmonic oscillator, with high accuracy and precision, considering  cubic and quartic perturbations. We intend to show that this system exhibits quantum-classical invariant chaotic behavior in the vicinity of vortices. For this, we vary the value of $\hbar$ and subsequently analyze the separation between the quantum trajectories. 
\section{de Broglie-Bohm  theory }
The de Broglie-Bohm  view of quantum mechanics is a non-local interpretation based on a classical analogy of Schrödinger equation, where the particle dynamics is driven by the guidance equations
\begin{align}\label{guidan}
   m\frac{d\mathbf{x}}{dt}=\hbar \,\mathrm{Im}\left\{ \frac{\nabla \Psi}{\Psi}\right\}.
\end{align}
In this perspective, the wave function guides the quantum particles in the configuration space. Assuming a polar form for the wave function $\Psi(\mathbf{x},t)=R(\mathbf{x},t)e^{iS(\mathbf{x},t)/\hbar}$ and substituting into Schrödinger equation, we obtain two real relations, namely
\begin{align}
   \partial_t S+\frac{(\nabla S )^2}{2m} + V + Q &=0,\label{Hamilton-Jacobi}\\
   \partial_t R^2 + \nabla\cdot\left( R^2\frac{\nabla S}{m}\right)&=0,\label{continuity}
\end{align}
with the quantum potential  $Q(\mathbf{x},t)$ defined as
\begin{align}
    Q\equiv-\frac{\hbar^2}{2m}\frac{\nabla^2R}{R}.
\end{align}
Eq. \eqref{Hamilton-Jacobi} is a Hamilton-Jacobi equation with an effective potential given by the sum of the classical and quantum potentials. Moreover, Eq. \eqref{continuity} is a continuity equation that reveals $|\Psi|^2=R^2$ as a probability density. Thus, given an initial distribution $|\Psi(\mathbf{x},0)|^2$, the trajectories probability density at any instant will be $|\Psi(\mathbf{x},t)|^2$. From Eqs. \eqref{guidan} and \eqref{Hamilton-Jacobi}, we obtain a quantum analogue of Newton's second law with an extra quantum force $-\nabla Q$, such that
\begin{align}
    m\frac{d^2\mathbf{x}}{dt^2}=-\mathbf{\nabla}V-\mathbf{\nabla}Q.\label{quantum force}
\end{align}
In this form, Eq. (\ref{quantum force}) is very analogous to the classical equations of motion, suggesting  the possibility of chaos. The main difference here is the presence of the quantum potential, which is, in general, nonlinear. Interestingly, some classical systems that do not exhibit chaotic behavior may present quantum chaos~\cite{Tzemos2020}. This happens because the trajectories obtained via guidance equations \eqref{guidan} undergo the influence of NPXPCs throughout their evolution ~\cite{Tzemos2020,Falsaperla2003,Conto2009}. The NPXPCs are formed by  two main elements: the nodal points, defined as regions where  $\mathrm{Re}{(\Psi)}=\mathrm{Im}{(\Psi)}=0$, and the X-points, which are unstable hyperbolic points defined in the reference frame of the nodal points that accompany its evolution. The complex geometry created by this pair generates quantum vortices, which are responsible for the scattering of  neighboring trajectories. So, if a particular system has a considerable number of such vortices, the nearby trajectories can experience significant deviation. Hence, it is natural to expect that such systems will exhibit chaotic behavior.
de Broglie-Bohm  theory 
\section{Model} 
Let us consider a  generic two-dimensional  anharmonic oscillator whose Hamiltonian is given as follows
\begin{align}
\label{Hamiltonian}
    \mathcal{H}=&\frac{p_x^2}{2m}+\frac{p_y^2}{2m}+\frac{1}{2}(\omega_x^2 x^2+\omega_y^2 y^2)\\\nonumber
    &-\kappa xy+\frac{\alpha}{3}(x^3+y^3)+\frac{\beta}{4}(x^4+y^4),
\end{align}
where $\alpha$ and $\beta > 0$ are constants of the cubic and quartic interactions, respectively. Without loss of generality, we set the mass $m$ and  frequencies as unity. The coupling constant $\kappa$ connects the two spatial coordinates. Setting $\kappa$ to zero results in two independent oscillators, while $\kappa\neq0$ results in a \textit{nonlinear membrane }.
The Eq. \eqref{Hamiltonian} under the condition $\alpha\gg\beta$ is highly unstable. To avoid this problem, we set $\alpha/3$ and $\beta/4$ to be small values, with $\alpha$ slightly greater than $\beta$. Here, the nonlinear interactions may be treated as small perturbations. 
\section{Methods} 
We numerically solve the time-dependent Schrödinger equation, with the Hamiltonian given by Eq. \eqref{Hamiltonian}, considering different values of $\kappa$ and $\hbar$. We employ the finite element method (FEM)~\cite{Bathe2008,WolframFEM1,WolframFEM2} in a square domain $\mathcal{D}\equiv [-L,L]\times[-L,L]$, where $L$ is the linear size of this domain, and we integrate the temporal part with high accuracy and precision, using adaptive time-steps $dt$ requiring a significant computational effort. For each  value of $\hbar$ we set $L$ in a region that the entire wave packet is included, but ensuring that this wave packet is far from the boundaries. The domain is discretized using uniform spacing square mesh elements, and Dirichlet boundary conditions are applied to prevent undesirable effects in the numerical solutions. The normalization is preserved within an error of $10^{-3}$. For numerical purposes, we consider as initial conditions a superposition of wave functions of the two-dimensional harmonic oscillator as follows:

\begin{equation}
    \Psi(\mathbf{x},0)=\frac{1}{2}\left(\psi_{00}(\mathbf{x})+\psi_{01}(\mathbf{x})+\psi_{10}(\mathbf{x})+\psi_{11}(\mathbf{x})\right),\label{IC}
\end{equation}
where the eigenvectors are given by the relation
\begin{equation}
    \psi_{nm}(\mathbf{x})=\dfrac{e^{-\frac{1}{2\hbar}(x^2+y^2)}}{\sqrt{ 2^{n+m}n!m!\pi\hbar}} H_{n}\left(\frac{x}{\sqrt{\hbar}}\right)H_{m}\left(\frac{y}{\sqrt{\hbar}}\right),
\end{equation}
with $H_{n}(x)$ the Hermite polynomials of order $n$. This choice of initial condition is interesting because it shares the same symmetry of the Hamiltonian \eqref{Hamiltonian}, which is invariant by changing  $x$ and $y$ variables. In other words, the line $y=x$ is an axis of symmetry of the system. By  Curie's causality \cite{Curie1894} we expect the manifestation of such symmetry within the calculation of the Bohmian trajectories. In our simulation, we choose $\alpha/3=0.05$ and $\beta/4=0.04$, ranging $\kappa$ from $0$ to $1$ and $\hbar$ from $0.05$ to $1$ at arbitrary steps. Eq.~\eqref{guidan} is integrated using 8th order Runge-Kutta method~\cite{Runge1895,Kutta1901,RugenKutta} with time-steps of $dt=10^{-5}$. 

In order to analyze the divergence of initially nearby trajectories, we compute their deviations $(\delta x,\delta y,\delta p_x,\delta p_y)$ in the fourth-dimensional phase space $(x,y,p_x,p_y)$, where $p_x$ and $p_y$ are simply the momentum components. Considering 60 different initial conditions, we perform the mean of  $\ln{\xi(t)}$, in which $\xi(t)$ are the normalized phase space distance of each pair of neighboring trajectories, defined as $\xi(t)=\tilde \xi(t)/\tilde \xi(0)$, where  $\tilde \xi (t)=\sqrt{\delta x(t)^2+\delta y(t)^2+\delta p_x(t)^2+\delta p_y(t)^2}$. For each pair we set initial space components equally distant by an amount of $\epsilon=10^{-4}$. 
\section{Results} 
\begin{figure}
    \centering
     \includegraphics[width=3.8cm]{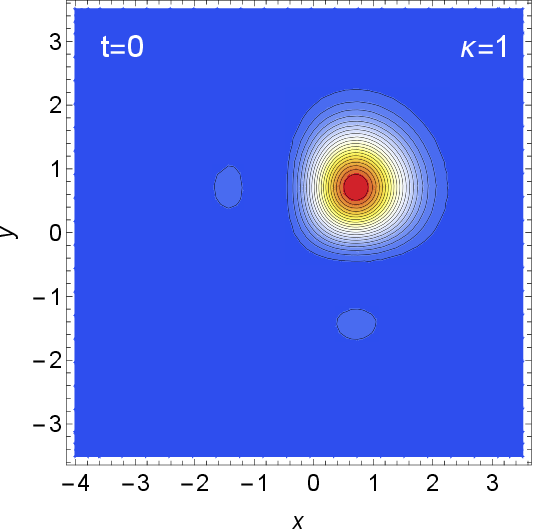} 
     \includegraphics[width=3.8cm]{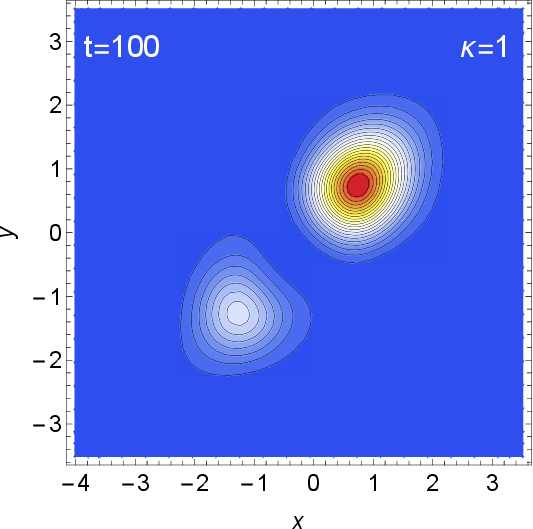}
     \includegraphics[width=0.8cm]{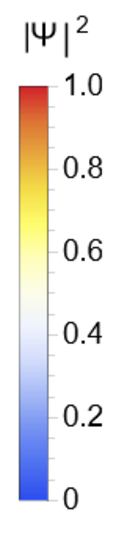}\\
     \includegraphics[width=3.8cm]{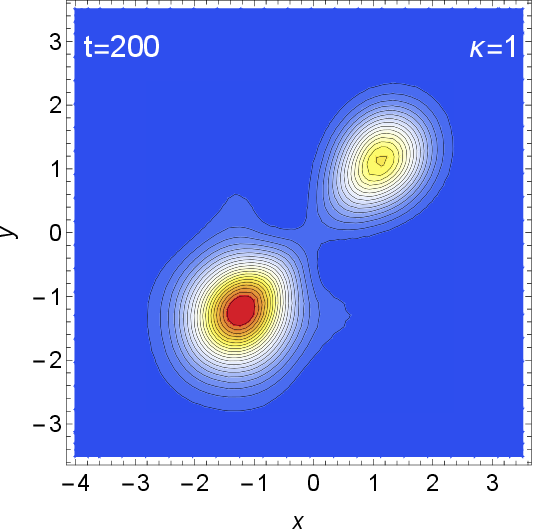} 
     \includegraphics[width=3.8cm]{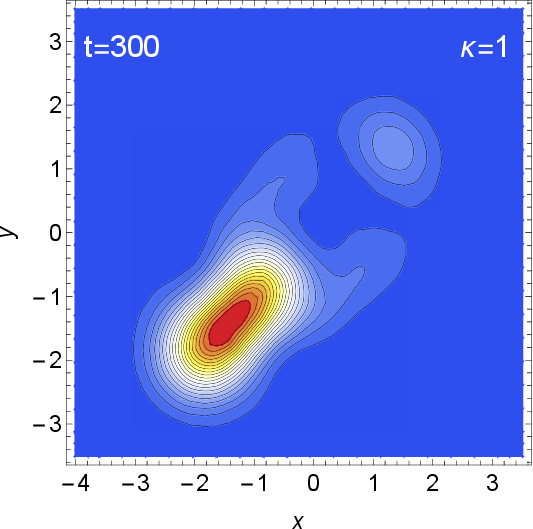}
     \includegraphics[width=0.8cm]{aaa.eps}      
        \caption{Evolution of the probability density $|\Psi(x,y,t)|^2$ of the generic anharmonic oscillator, for $\kappa=1$, at (top left) $t=0$, (top right) $t=100$, (bottom left) $t=200$ and (bottom right) $t=300$.}
    \label{wavefunction}
\end{figure}

\begin{figure}
    \centering
     \includegraphics[width=5.cm]{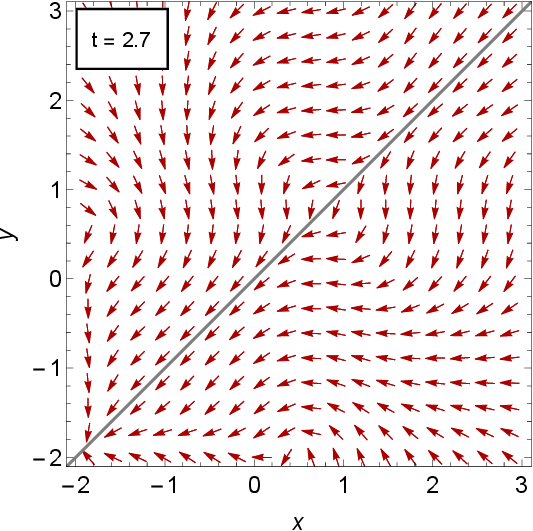}  
     \includegraphics[width=5.cm]{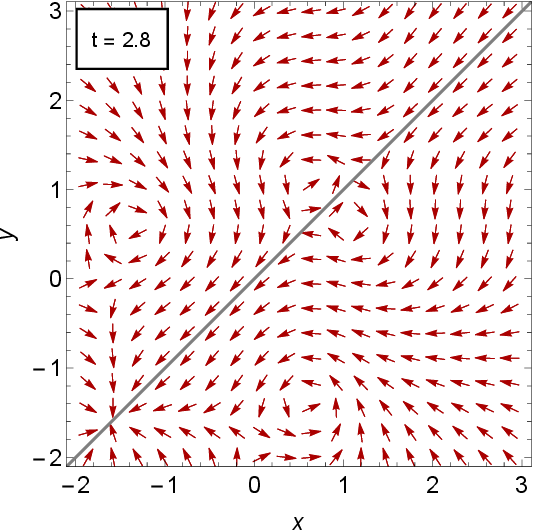}  
     \includegraphics[width=5.cm]{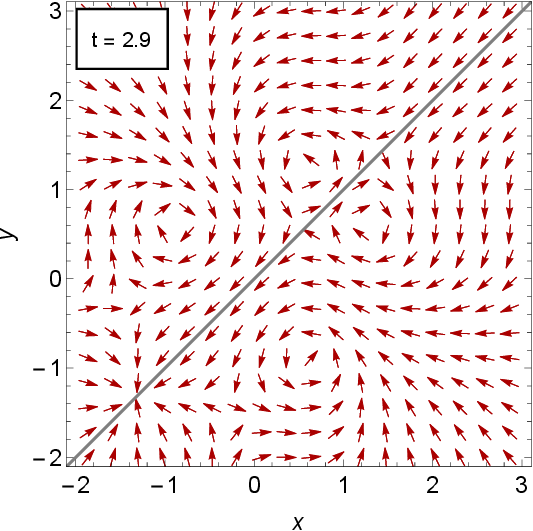}  \\
     \includegraphics[width=5.cm]{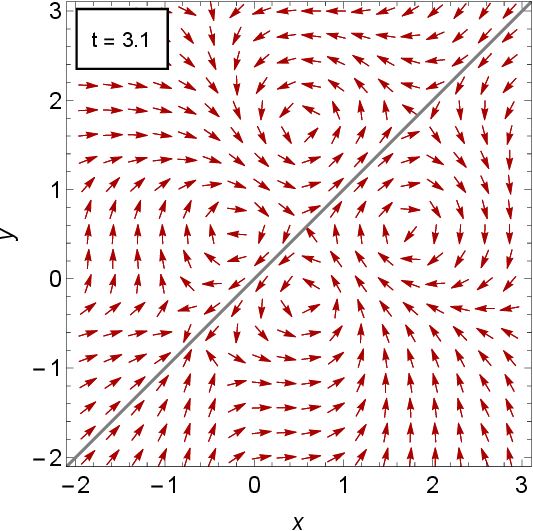}
     \includegraphics[width=5.cm]{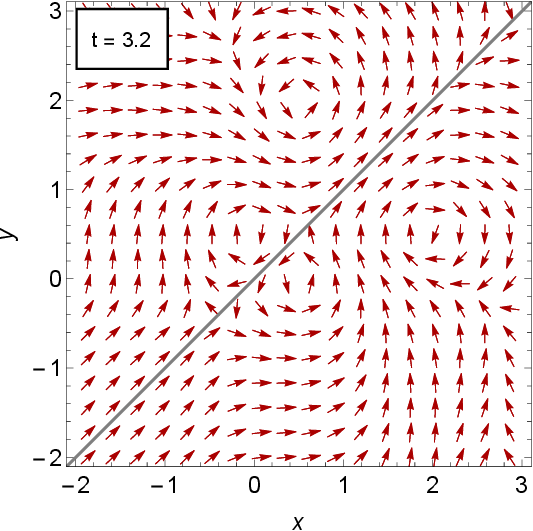}
     \includegraphics[width=5.cm]{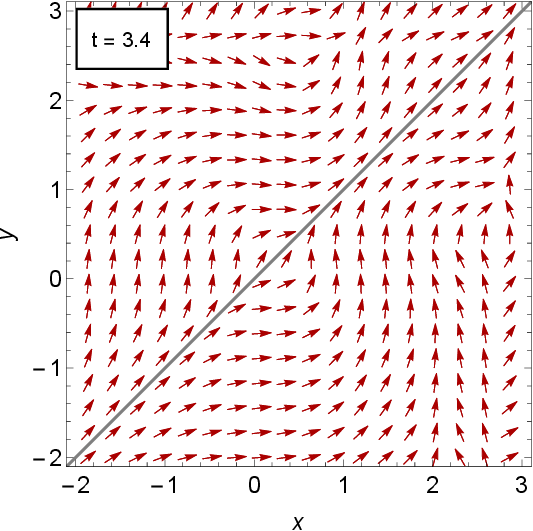}
        \caption{Ilustration of the emergence of  four dynamical vortices in a square region $\mathbf{x} \in [-2,3]\times [-2,3]$ at  $t=2.7,2.8,2.9,3.1,3.2$ and $t=3.4$ for $\kappa=0.1$. We choose $\kappa=0.1$ for didactic purposes (number of vortices increase as $\kappa$ also increases). }
    \label{vortex}
\end{figure}

Let us present now the results of our simulations. We notice that the wave packet spreads during its evolution, as shown in Fig. \ref{wavefunction}. This spreading is due to the $-\kappa xy$ interaction and the unstable cubic potential, which assumes both positive and negative values. When $|x|,|y|>1$, the cubic potential becomes very unstable, leading the system to slightly escape from the potential well. However, the quartic interaction in Eq. \eqref{Hamiltonian} is responsible for stabilizing the oscillations in such a way that neither $-\kappa xy$ nor the cubic term cause an exaggerated stretching of the wave function.
Analyzing the solutions of Eq. \eqref{guidan}, we illustrate the velocity field in the case $\kappa=0.1$ (see Fig.~\ref{vortex}). Within a short period of time, we observe the presence of four dynamic vortices, where two distinct events can be highlighted: Firstly, we notice the formation of a vortex pair at $t=2.8$ and, right after, at $t=3.2$, we detect the approximation and eventual collision of another pair, resulting in their disappearance. Furthermore, dynamic vortices continue to emerge in the same area and undergo similar collision phenomena. This is a direct effect of the previously mentioned symmetry. Since the Hamiltonian \eqref{Hamiltonian} and the initial condition \eqref{IC} are invariant under the exchange of the $x$ and $y$ variables, the equations of motion should be invariant under the reflection in relation to the line $y=x$. Hence, the vortices are created and annihilated in pairs equally distant of this axis of symmetry, having the same absolute value of vorticity ($\nabla \times \mathbf{v}$), but with opposite signs. As a result, the pairs have the same diameter but spinning in opposite directions. It is worth to emphasize that the production of vortices is directly related to the increase of the coupling constant $\kappa$. 
\begin{figure}
    \centering           
    \includegraphics[width=6.5cm]{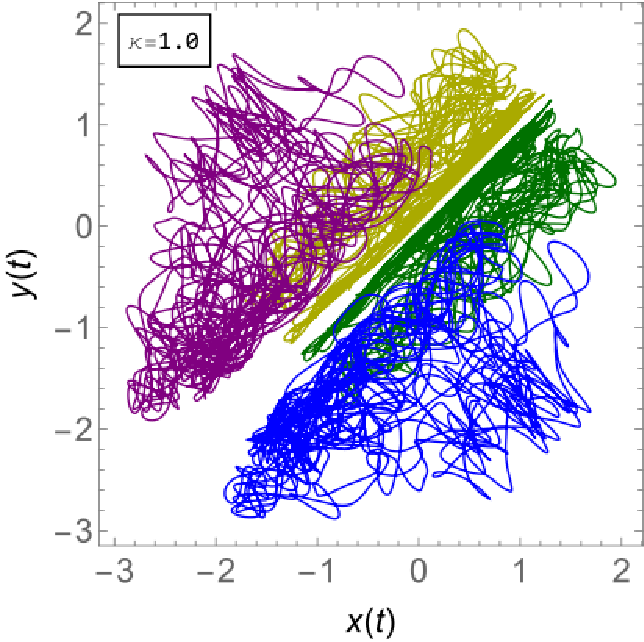}
    \includegraphics[width=6.5cm]{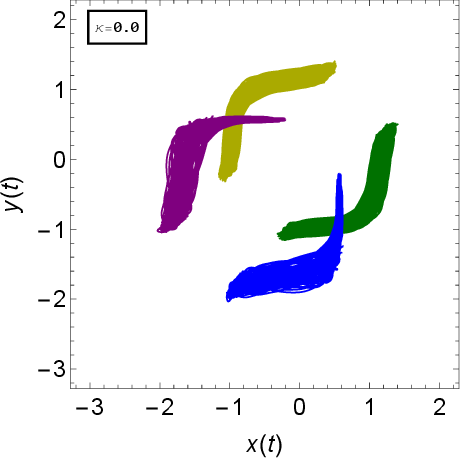}
        \caption{\justifying Parametric plot of the quantum trajectories $(x(t),y(t))$ considering initial conditions near vortices collision and repulsion. {\textit{ Left:}} $(x_0,y_0)=\{(1.4,0.5),(0.5,1.4),(0.6,-0.5),(-0.5,0.6)\}$  with $\kappa=1$. \textit{Right:}  with $\kappa=0$ and the same previous initial conditions. Note the reflection symmetry with respect to the line $y=x$.}
    \label{chaos1}
\end{figure}

In Fig. \ref{chaos1} we show the parametric plot of the guidance equations solution, for different initial conditions. It is possible to observe that all four trajectories present very distinct and seemingly unpredictable behaviors when $\kappa=1.0$. In contrast, the trajectories of the $\kappa=0$ case are clearly ordered. This indicates that $\kappa\neq0$ is crucial for inducing  chaotic behavior. It is noteworthy that despite the inherent nonlinearity caused by the quantum potential, it alone is not sufficient for chaos to emerge. Without any constraint or coupling  to bound the spatial coordinates, the emergence of unpredictability and sensitivity to initial conditions are nearly null, because there are no sufficient vortices to induce chaotic behavior and  the dynamics are not conducive to chaos as well. Additionally, we  notice that the same symmetry regarding the exchange of the spatial variables also occurs in the level of the quantum trajectories, supporting the effectiveness of our numerical results. 

\begin{figure}
    \centering    
    \includegraphics[width=6.5cm]{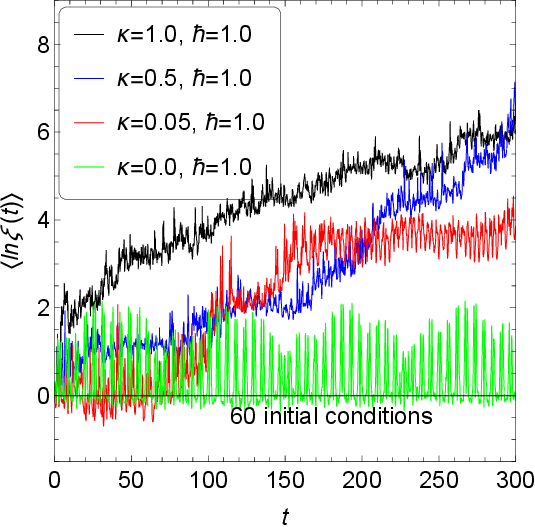}
    \includegraphics[width=6.3cm]{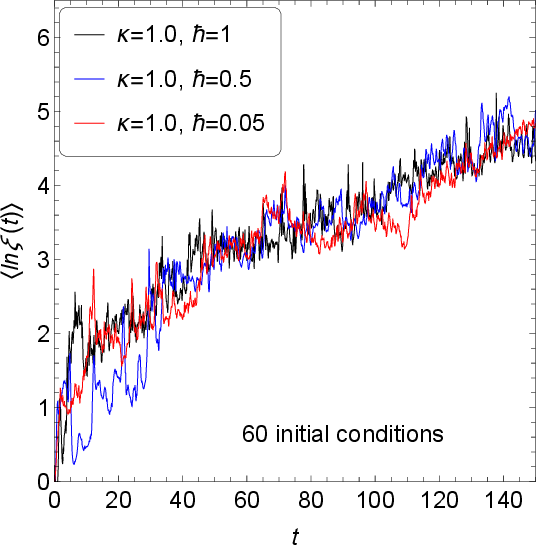}
            \caption{{\it Left:} $\langle \ln{\xi(t)} \rangle $ versus $t$ for $\hbar=1$ and $\kappa=0,0.05,0.5,1$ with 60 pairs of initial conditions regularly spaced in the  interval of $(x_0,y_0)\in [-1.5,-1.1]\cup[1.1,1.5]$. {\it Right:} $\langle \ln{\xi(t)} \rangle $ versus $t$ for $\kappa=1$ and $\hbar=0.05,0.5,1$ with 60 pairs of initial conditions regularly spaced in the  interval of $(x_0,y_0)\in[-0.5,-0.1]\cup[0.1,0.5]$.}
    \label{sensitive}
\end{figure}
Another characteristic of chaos is the sensitivity to initial conditions, where small perturbations in the system can result in  significantly  different outcomes. In Fig.~\ref{sensitive}, we notice that $\langle \ln{\xi(t)\rangle}$ is a nearly monotonically increasing function of time, when $\kappa>0$, indicating an exponential deviation of the trajectories. Conversely, when $\kappa=0$, no evidence exists of an exponential deviation, indicating a regular behavior.

Also analysing the results from Fig. \ref{sensitive}, $\hbar$ has almost  no influence in the deviation of the trajectories. Indeed,  the same angular coefficient is found varying $\hbar$ from $0.05$ to $1$. In other words, the system studied here presents chaotic behavior in both classical and quantum regimes. Therefore, through the Bohmian approach of quantum systems, it is possible to study the quantum chaotic dynamics employing the same techniques used to study classical chaos, resulting in robust chaos in both limits ($\hbar=1$ and $\hbar\to 0$), showing that quantum and classical chaos can be seen as two faces of the same coin.
\section{Conclusions}
Summarizing, we  simulate, within the Bohmian quantum mechanical approach, a two-dimensional anharmonic oscillator under the influence of  a coupling potential and of a tunnable Planck constant.  The generic requirements for chaos (unpredictability and sensitivity to small perturbations of the initial conditions)
are satisfied by this system. In the absence of coupling (i.e., $\kappa \to 0$), chaos disappears. In remarkable contrast, no important influence is observed along the quantum-to-classical crossover ($\hbar \to 0$), thus validating the conjecture that no strong difference exists between quantum and classical chaos.
Couplings  as in Eq. \eqref{Hamiltonian} are  good candidates for inducing  quantum chaos (both strong and weak \cite{TsallisLloyd}) and possibly enlighten the study of emergence of vortices in quantum systems.  

We acknowledge  Nelson Pinto-Neto for fruitful discussions as well as CNPq and FAPERJ (Brazilian agencies) for  partial financial support .

\end{document}